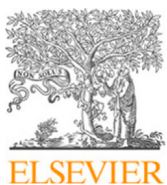
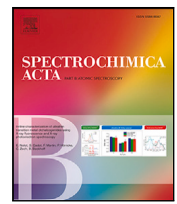

# Comparison of two laser wavelengths for LIBS bioimaging of plants grown in lunar regolith☆

Tomáš Vozár [a], Ludmila Čechová [a],*, Jakub Buday [a,b], Martin Füleky [c], Katarína Molnárová [c], Libor Lenža [c], Jiří Mužík [d], Yuya Koshiba [d], Martin Smrž [d], Pavel Pořízka [a,b], Jozef Kaiser [a,b]

[a] *Central European Institute of Technology (CEITEC), Brno University of Technology, Purkyňova 123, Brno, 612 00, Czech Republic*
[b] *Faculty of Mechanical Engineering, Brno University of Technology, Technická 2896, Brno, 625 00, Czech Republic*
[c] *Department of Chemistry and Biochemistry, Mendel University in Brno, Zemědělská 1665/1, Brno, 613 00, Czech Republic*
[d] *HiLASE Centre, Institute of Physics of the Czech Academy of Sciences, Za Radnicí 828, Dolní Břežany, 252 41, Czech Republic*

ABSTRACT

The colonisation of extraterrestrial planets requires sustainable food production independent of Earth-based supplies. Due to the high costs and complicated logistics of food transport, in-situ cultivation will be essential. Growing plants directly in regolith offers a practical approach to achieve sustainable long-term human habitation beyond Earth. In this study, Laser-Induced Breakdown Spectroscopy (LIBS) technique was employed for bioimaging of broccoli (*Brassica oleracea*) and salad (*Lactuca sativa*) plants grown in Lunar regolith simulant and control substrate. For this purpose, the potential of the 2090 nm laser wavelength for bioimaging of plant tissue was studied compared to the conventional 1064 nm. The signal-to-noise ratio (SNR), total emissivity ($\varepsilon_{tot}$), and Mg II / Mg I intensity ratio (ionisation degree) were all higher when using the 2090 nm laser wavelength compared to 1064 nm. These findings indicate that the 2090 nm laser produces a hotter and more efficiently ionised plasma, supporting its feasibility for bioimaging of plant tissues. Additionally, bioimaging with both laser wavelengths has confirmed higher uptake of key plant nutrients such as magnesium (Mg) and calcium (Ca) from Lunar regolith simulant. This results support the potential of LIBS as a diagnostic tool for plant growth monitoring in extraterrestrial environments.

## 1. Introduction

The colonisation of extraterrestrial bodies stands as one of humanity's greatest challenges in the 21st century. One of the crucial problems of wider space exploration and colonisation is providing food supplies. Transporting food from Earth is unsustainable due to logistical problems and high travel costs. Therefore, establishing food production on-site will be essential for long-term colonisation efforts. One of the most straightforward approaches is to grow plants directly on the extraterrestrial surface in regolith [1,2].

Among the most promising candidates for colonisation are Mars and the Moon, due to their relative proximity to Earth and the extensive studies of their surfaces and atmospheres. Both are covered by a layer of extraterrestrial soil known as regolith [1–3]. Unlike fertile soil on Earth, regolith poses challenges for plant growth due to its compact structure and composition, which complicate root development and plant growth. Although regolith contains many essential elements for plants, nitrogen is missing [3], an essential nutrient required for almost all plant growth [4]. On Earth, nitrogen is supplied primarily by organic matter, which is absent in the lunar regolith [4–9]. Many essential elements are not present in readily accessible chemical forms for plants [9]. In addition, certain toxic elements could further hinder plant development. For example, aluminium and chromium are considered potentially toxic and growth limiting [10–13]. Regolith also contains iron and manganese, which are essential plant nutrients. However, when present in excessive concentrations, as it is often the case in regolith, they can become toxic and negatively affect plant growth [14–16].

Despite these challenges, recent studies have demonstrated that plants are capable of germinating and growing in regolith, highlighting their potential as a substrate for future space agriculture [2,17]. For instance, in 2022, Paul et al. investigated the growth of *Arabidopsis thaliana* in lunar regolith samples collected during the Apollo 11, 12, and 17 missions. Their findings showed that while *A. thaliana* germinated and grew successfully in various types of lunar regolith, the plants developed more and exhibited several stress-related morphologies, likely reflecting the chemical limitations and toxicity of the substrate [2].

Currently, several approaches are being explored to mitigate the hostile properties of regolith and enhance its suitability for plant growth. Lehner et al. (2018) demonstrated that microbes can be used to remove the effects of heavy metals and increase the concentration of beneficial biomolecules in regolith, thus improving its potential as a growth substrate [18]. Similarly, Duri et al. (2020) attempted to improve the fertility of Martian regolith simulants by incorporating compost as a source of organic matter. The addition of compost not only increased the yield of the plants, but also improved overall quality [19]. Wang et al. (2025) explored the use of organic waste processed by earthworms as a means to enrich the regolith and improve its suitability for plant growth. In addition, earthworms improved porosity, ion adsorption capacity, and significantly mitigated compaction and salinisation of Lunar regolith simulant [20]. The study by Caporale et al. (2022) investigated the mixing of Lunar and Martian regolith simulants with commercial horse and swine manure. This amendment significantly enhanced both above- and below-ground plant biomass. Another important factor to consider is the sharpness of Lunar soil particles, which can damage plant roots [21]. Yao et al. (2022) proposed bioweathering as a strategy to address this issue, both by reducing particle sharpness to prevent root injury at the micro level and by enhancing substrate porosity at the macro level. Their results showed that bioweathering significantly improved the properties of Lunar regolith simulant for plant growth [22].

Typically, elemental analysis of whole plants or their parts is carried out using wet chemistry methods, for example, ICP-OES (Inductively Coupled Plasma Optical Emission Spectroscopy), following acid digestion of the plant material [23]. Even though these techniques are accurate and sensitive, they suffer from significant drawbacks, such as sample preparation that is both time-consuming and expensive. Laser-Induced Breakdown Spectroscopy (LIBS) is an advanced analytical technique used for rapid, in-situ elemental analysis. A laser pulse of high fluence is focused on the sample, providing enough energy to heat, evaporate, and ionise the material, which results in plasma formation. As the plasma cools down, excited atoms and ions emit light at characteristic wavelengths [24]. The ability to perform rapid, multi-element analysis directly from solid samples makes LIBS a highly versatile tool for plant nutrition diagnosis [25,26]. An important advantage of LIBS compared to wet chemistry techniques is its ability to preserve spatial information, enabling direct analysis of the distribution of individual elements across different tissues or even within cells [27,28].

LIBS has already been utilised to map the distribution of different nanoparticles or toxic/heavy metals in plant organs. Also, it localises the exact places where these nanoparticles or toxic/heavy metals are most likely stored when growing plants in contaminated soils [29]. Thanks to the ability of spatial resolution measurement by LIBS, it is possible to localise the exact parts of the plants where the most probable location of storage is. Since regolith could also contain toxic elements, it is important to detect their presence in plants. Furthermore, LIBS can also visualise the uptake, transport, and spatial distribution of nutrients derived from substrates [28,30]. In most of these applications, bioimaging is carried out using a 1064 nm laser and its harmonics (532 nm and 266 nm) [28,29,31]. The plant matrix is composed mainly of organic compounds and biochemicals. These molecules account for many overlapping absorption features in the Short-Wave-infrared region (SWIR, 1500–2000 nm) compared to the near-infrared (NIR, 700–1000 nm). Although the use of longer wavelengths is less common, the higher absorbance in SWIR could open new opportunities and enhance detection limits in biological matrices, such as plants [32].

The novelty of this study is two-fold: it demonstrates the feasibility of a 2090 nm laser for LIBS plant tissue bioimaging and provides new insights into plant elemental uptake from Lunar regolith simulant.

The first task of this study was to employ Laser-Induced Breakdown Spectroscopy using two different laser wavelengths for plant bioimaging and to evaluate the effectiveness of a 2090 nm laser compared with a 1064 nm laser for the elemental analysis of plant tissues. Due to the higher absorption of 2090 nm laser light by water and organic compounds in plant tissues compared to 1064 nm, the laser–matter interaction at this wavelength is expected to be more efficient. This could make the 2090 nm wavelength more suitable for ablating biological samples and enhancement of the detection of trace elements in plants [33,34]. The following task aims to provide new insights into how plants interact with lunar regolith simulant during growth. Salad (*Lactuca sativa*) and Broccoli (*Brassica oleracea*) plants were grown in Lunar regolith simulant (LMS-1, Exolith Lab, USA) and a control growth substrate for one month. We employed LIBS to monitor the uptake of elements from the regolith by plants, including both nutrients and potentially toxic components. These results could advance the development of sustainable in-situ plant production beyond Earth and support the use of LIBS as a diagnostic tool for plant growth monitoring in sustainable space-agricultural systems.

## 2. Materials and methods

### 2.1. Plant cultivation and sample preparation for LIBS analysis

Salad (*Lactuca sativa*) and Broccoli (*Brassica oleracea*) plants were cultivated for one month in three substrate treatments—Lunar regolith simulant LMS-1 (Exolith Lab, USA), control substrate (Plagron Growmix, 50l), and control substrate supplemented with fertiliser (Terra Aquatica TriPart, France). The composition of the LMS-1 substrate was previously investigated by Isachenkov et al. (2022), who reported on its elemental and chemical characteristics. LMS-1 is characterised by an inorganic composition dominated by silicate and metal oxides, primarily $SiO_2$, FeO, $Al_2O_3$, CaO, and MgO [35,36]. Plagron Growmix is primarily an organic substrate, based on peat and worm castings, and is pre-fertilised with a mineral NPK fertiliser (216–252–432 $g\,m^{-3}$). The control substrate supplemented with fertiliser was the same Plagron Growmix; however, during plant growth, it was additionally supplied with TriPart fertiliser. TriPart is a three-component fertiliser, with each component tailored to a specific vegetative stage of plant growth. It provides both macronutrients (e.g. N, P, K, Ca and Mg) and micronutrients (e.g. Fe, Mn, B, Mo) essential for optimal plant growth. The plants were grown under controlled conditions of 16 h light/8 h dark photoperiod, $23\,°C$, and a light intensity of 100 $\mu mol\,m^{-2}\,s^{-1}$. Both control and LMS-1 treatments were watered with $ddH_2O$ (dd = double distilled). At harvest, leaves were carefully removed from pots, pressed to maximise flatness, and dried. Dried leaves were then fixed on microscope slides using epoxy resin. Thus, prepared plant samples were fixed onto the microscopy slides using epoxy resin. The methodology for plant analysis by LIBS was done according to previously published work by Modlitbová et al. (2020) [37].

### 2.2. LIBS plant bioimaging

Plant bioimaging was performed using Laser-Induced Breakdown Spectroscopy. The 2D elemental maps were recorded using two different LIBS ablation wavelengths. The first customised cage LIBS system consisted of a Nd:YAG laser with a 1064 nm ablation wavelength (8.8 mJ, 20 Hz, 5.5 ns, Litron). The second customised cage LIBS system consisted of a Ho:YAG laser with a 2090 nm ablation wavelength (2.9 mJ on sample, 100 Hz, 4.1 ns, HiLASE Centre, Czech Republic [38]). Pulse-to-pulse energy fluctuations are in the range of below 1% for both lasers. The laser energies were carefully adjusted based on irradiance ($I$) calculations in order to ensure equivalent ablation conditions for both lasers (1). Where $t$ is pulse duration, $S$ is the spot area and $E$ is the energy of the laser. The diameter of the spot was measured on steel, with a 10 μm diameter for the 2090 nm laser and 15 μm for the 1064 nm.

$$I = \frac{E}{t \cdot S} \quad (1)$$

Bioimaging was done with a step size of 40 μm in the case of a 1064 nm laser and 20 μm in the case of a 2090 nm laser under ambient atmospheric conditions. The laser beam was directed into the interaction chamber by a series of laser-line mirrors and subsequently focused onto the sample using a doublet lens. For the 2090 nm wavelength, an E-coated 50 mm lens was used, while for the 1064 nm wavelength, a 32 mm triplet lens was employed. Emission of the plasma plume formed by the ablation of material was collected using a UV-graded collection lens (75 mm) and coupled into an optical fibre with a 400 μm core diameter. Emitted light was transmitted to a Czerny–Turner spectrometer (Avantes, AvaSpec) with a CMOS detector and a wavelength range of 247–410 nm (resolution 0.042 nm), with an exposure time of 50 μs and a gate delay of 1 μs. The same spectrometer and collection optics was used in both laser systems to preserve the same conditions in comparative measurements.

Data obtained from elemental bioimaging were processed using a custom Python script. The datasets were first corrected for background using the moving minimum method. The background-corrected data were then used for subsequent analyses. Spectral lines selected for 2D elemental mapping included Mg I 285.20 nm, Mg II 279.50 nm, Ca II 393.36 nm, and Ca II 396.75 nm. For each elemental line, a signal-to-noise ratio (SNR) map was calculated by dividing the line intensity by the standard deviation of the noise region. Only data corresponding to plant tissue were included; signals originating from the epoxy resin were excluded. To achieve this, SNR maps were masked by selecting manually by polygon selector only spectra originating from plant tissue ablation, thereby omitting spectra from the epoxy resin.

For the comparison of plasma properties at different ablation wavelengths, the total plasma emissivity ($\varepsilon_{tot}$) (2) was calculated by integrating the background corrected spectral emission across the 247–410 nm range.

$$\varepsilon_{tot} = \int_{\lambda_1}^{\lambda_2} I(\lambda) \, d\lambda \quad (2)$$

The plasma temperature can be correlated to the intensity ratio of Mg II/Mg I [39]. The range of our spectrum is from 247 to 410 nm; thus, for this purpose, the ionised Mg II 279.50 nm line and the atomic Mg I 285.31 nm line were chosen. The correlation is grounded in the principles of local thermodynamic equilibrium (LTE), where the populations of atomic and ionic species follow the Saha and Boltzmann distributions. The Saha equation describes the ionisation equilibrium between Mg I and Mg II, indicating that the ratio of ionised to neutral species increases exponentially with temperature (3) [39]. The Boltzmann distribution governs the population of excited states within these species, affecting the emission intensities of their spectral lines.

$$\frac{n_{Mg\,II}}{n_{Mg\,I}} = \frac{2}{n_e \Lambda^3} \frac{g_{Mg\,II}}{g_{Mg\,I}} \exp\left(-\frac{E_{ion}}{k_B T_e}\right) \quad (3)$$

In this equation, $n_{Mg\,II}$ and $n_{Mg\,I}$ represent the number densities of singly ionised and neutral magnesium atoms. $n_e$ is the electron density, $\Lambda$ is the thermal de Broglie wavelength, $g_{Mg\,II}$ and $g_{Mg\,I}$ are the statistical weights of the ionic and atomic ground states, $E_{ion}$ is the first ionisation energy of magnesium, $k_B$ is the Boltzmann constant, and $T_e$ is the electron temperature.

## 3. Results and discussion

### 3.1. Two ablation wavelength comparision for bioimaging

The LIBS setups for bioimaging consisted of two different ablation wavelengths, 2090 nm and 1064 nm. To consider the effectiveness of each wavelength for the ablation of plant tissue, we compared plasma characteristics, including the signal-to-noise ratio (SNR), total emissivity, and Mg II/Mg I ratio (ionisation degree). By evaluating these parameters, the impact of the laser wavelength on bioimaging performance was assessed.

**Table 1**
Mean SNR and Mg II / Mg I ratio for both laser wavelengths.

| Parameter   | 2090 nm | 1064 nm |
|-------------|---------|---------|
| SNR Mg I    | 35.12   | 23.22   |
| SNR Ca II   | 166.09  | 120.67  |
| Mg II / Mg I| 1.96    | 1.76    |

Fig. 1 presents a representative spectrum of salad (*Lactuca sativa*) control plant obtained by both ablation wavelengths. Both spectra show strong signals of nutrient elements such as singly ionised Mg II (279.65 and 280.27 nm), the atomic line of Mg I 285.31 nm, and singly ionised Ca II (315.89; 318.13; 370.60; 373.69; 393.37, and 396.85 nm). In the spectrum, there is also a weak signal of atomic lines of Si I 288.16 nm, Al I (308.21 and 309.27 nm), and singly ionised Ti II multiplet lines around 334 nm.

Fig. 2 presents signal-to-noise ratio (SNR) heatmaps of fertilised control salad (*Lactuca sativa*) plants. Salad leaf was ablated by two different laser ablation wavelengths (2090 nm on the left side and 1064 nm on the right side). The colour scale is the same for both wavelengths to ensure comparability between both two wavelengths. The heatmap shows atomic Mg I 285.31 nm and singly ionised Ca II 393.37 nm. These two lines were chosen to compare the effectiveness of the two ablation wavelengths as important nutrient elements. The heatmap of Mg shows a stronger signal when ablated by the 2090 nm laser wavelength compared to the 1064 nm laser wavelength. Table 1 shows that the mean SNR inside the mask for Mg was higher when using the 2090 nm laser compared to the SNR value obtained with the 1064 nm laser. Similarly, the mean SNR for Ca was higher with the 2090 nm laser than with the 1064 nm laser.

Fig. 3 presents total emissivity heatmaps ($\varepsilon_{tot}$) for both ablation wavelengths on the right side and Mg II / Mg I ratios (ionisation degree) on the left side. The colour scale is the same for both wavelengths to ensure comparability between both wavelengths. For the ratio calculation, ionised Mg II 279.50 nm and Mg I 285.31 nm atomic lines were used. The heatmaps clearly show that using a 2090 nm ablation wavelength results in higher emissivity compared to 1064 nm. This indicates that the plasma generated by 2090 nm ablation emits more strongly than that produced at 1064 nm. Also, the mean value of the Mg II / Mg I ratio inside the mask for the 2090 nm laser was higher compared to the ratio obtained with the 1064 nm laser (Table 1). Since the Mg II / Mg I ratio increases with plasma temperature, this suggests that the plasma generated by the 2090 nm ablation had a higher temperature than that generated by the 1064 nm ablation [39]. As previously mentioned, compared to 1064 nm, absorption of the 2090 nm wavelength by the plant matrix is higher [32], which can improve laser–matter interaction, leading to more efficient plasma formation. This may result in the hotter generated plasma and the improved signal intensity.

Peng et al. (2018) compared 532 nm and 1064 nm Nd:YAG lasers for the detection of chromium in rice (*Oryza sativa*) leaves using LIBS. They found that at lower pulse energies (< 40 mJ), excitation at 532 nm produced stronger spectral emission and yielded higher signal-to-noise (SNR) and signal-to-background ratios (SBR) than 1064 nm [40]. De Carvalho et al. (2015) compared laser excitation at 1064, 532, and 266 nm for LIBS analysis of pelletised plant material. They showed that shorter wavelengths produced better signal quality owing to improved laser–matter coupling, whereas the longer 1064 nm wavelength generated larger craters and more variable plasmas, resulting in reduced signal quality [41]. While Laser-Induced Breakdown Spectroscopy (LIBS) using conventional 1064 nm Nd:YAG lasers or their shorter harmonics is well-established for plant bioimaging [28,29,37,40,42,43], the use of longer-wavelength sources, such as 2090 nm Ho:YAG lasers, remains largely unexplored. These results support the feasibility of using 2090 nm for plant bioimaging and nutrient detection.

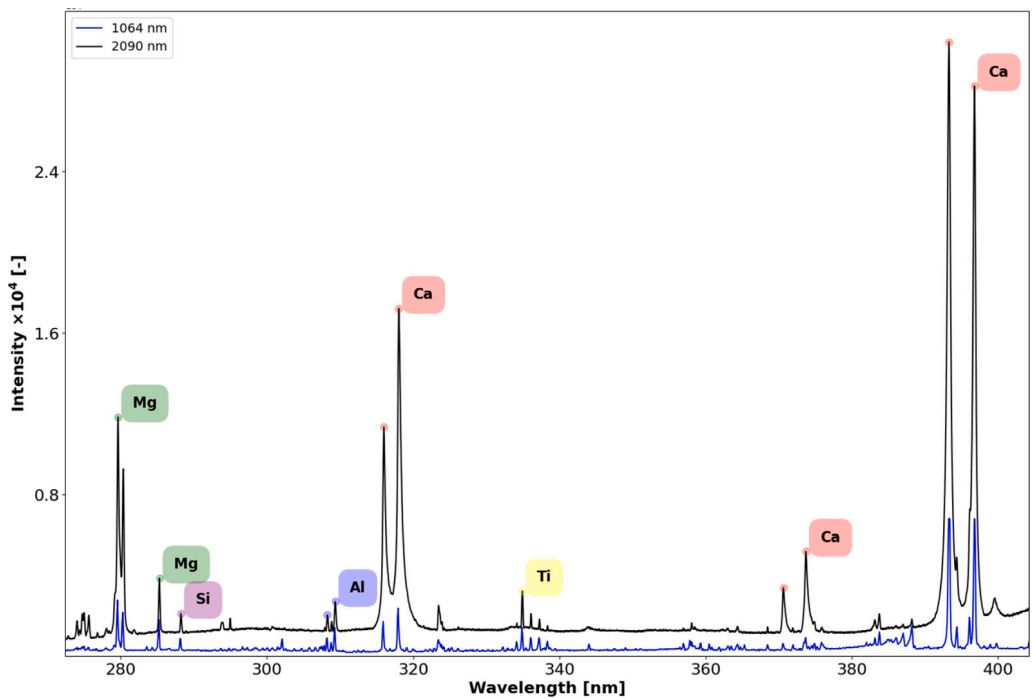

**Fig. 1.** The maximum LIBS spectrum of salad (*Lactuca sativa*) control plant ablated with two different ablation wavelengths. The blue spectrum corresponds to the sample ablated with a 1064 nm laser, whereas the black spectrum corresponds to ablation with a 2090 nm laser.

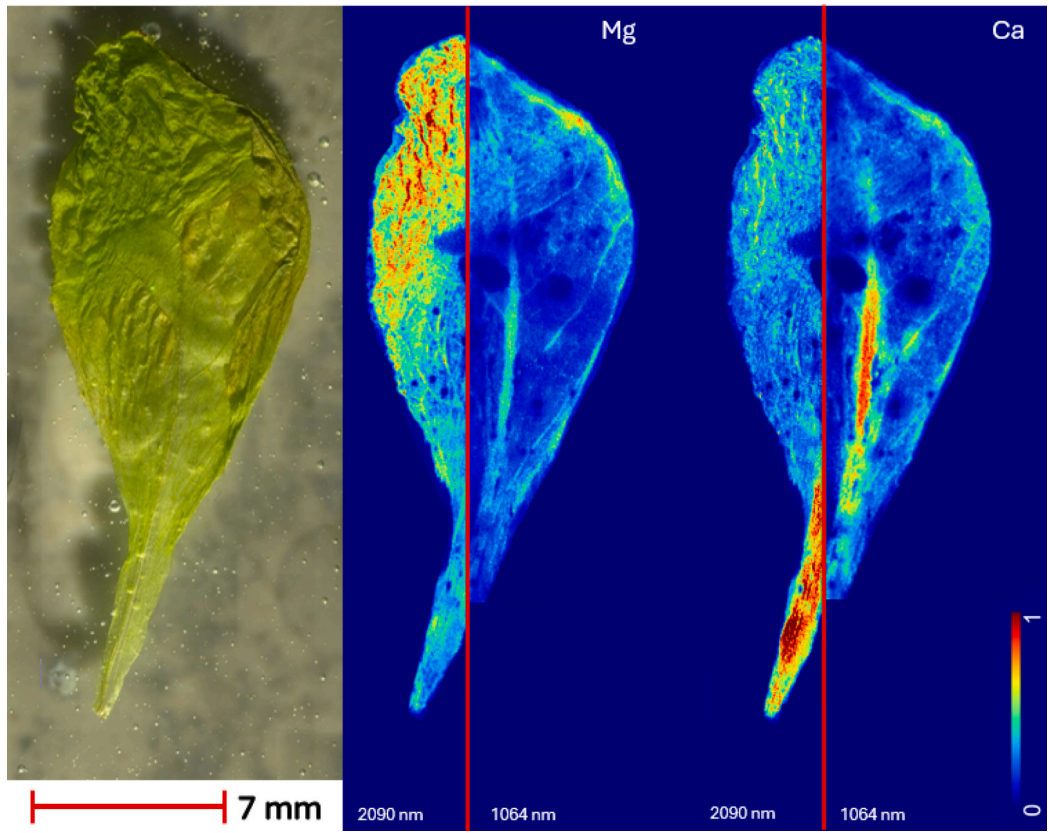

**Fig. 2.** Signal-to-noise ratio (SNR) heatmaps of fertilised control salad plant leaf ablated using 1064 nm and 2090 nm laser wavelengths. The heatmap is showing atomic Mg I 285.31 nm and singly ionised Ca II 393.37 nm. The colour scale is the same for each element and for both laser ablation wavelengths.

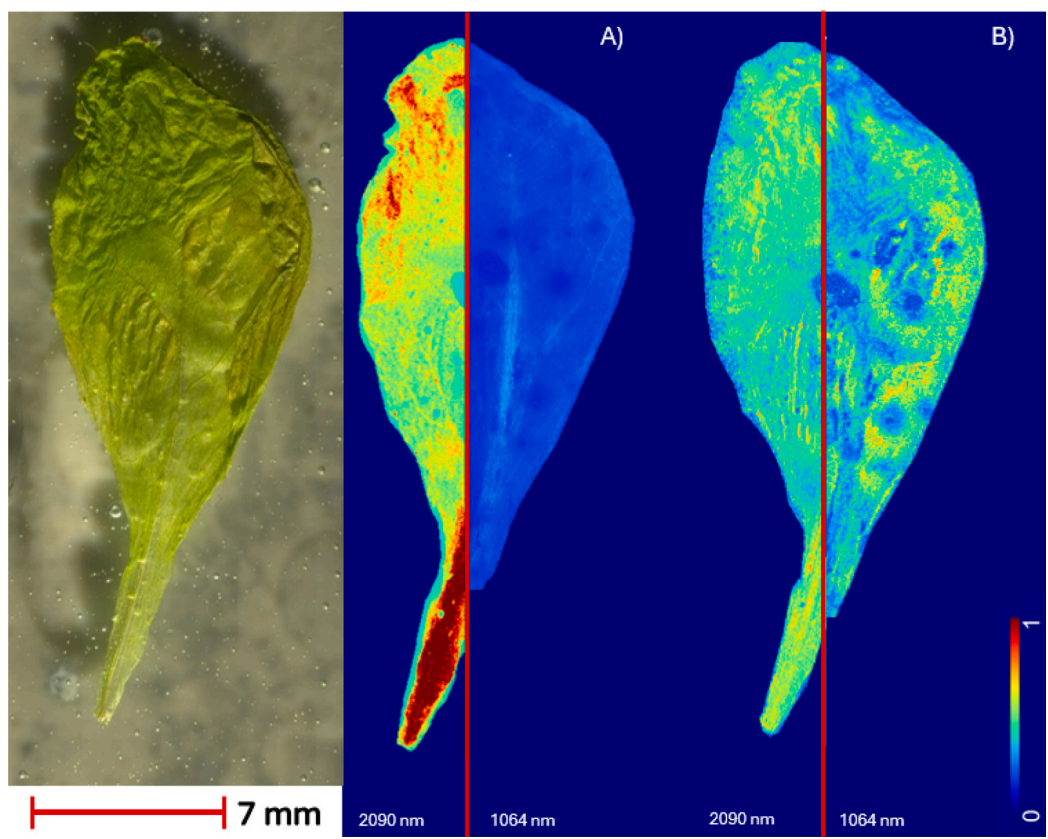

**Fig. 3.** Total emissivity ($\varepsilon_{tot}$) heatmap of a fertilised control salad plant leaf (A), and the Mg II / Mg I intensity ratio (ionisation degree) heatmap (B). Both heatmaps share the same colour scale for direct comparison of the two laser ablation wavelengths.

*3.2. Lunar regolith simulant plants bioimaging*

Bioimaging was performed using both laser ablation wavelengths, 2090 nm and 1064 nm. Analysis was used to determine the nutrient uptake changes in plants that were grown in Lunar regolith simulant compared to the control substrate. After data reconstruction, 2D SNR maps were obtained. As LIBS allows us to detect more elements at once, we are able to detect nutrient distribution in plants. Also, SNR intensities in these maps provide us with relative content information. As representative important plant nutrients, Magnesium (Mg) and Calcium (Ca) were chosen.

Fig. 4 shows 2D elemental SNR maps of Broccoli (*Brassica oleracea*) obtained by both ablation wavelengths. The colour scale is the same in both variants within the same ablation wavelength, to ensure visual comparability between variants. The maps show broccoli plants grown in the control substrate and the Lunar regolith simulant. For the maps, atomic Mg I 285.31 nm and singly ionised Ca II 393.37 nm emission lines were chosen. The maps show visibly increased amounts of magnesium and calcium accumulated in plants that were grown in Lunar regolith simulant compared to control plants.

Similarly to broccoli, salad (*Lactuca sativa*) plants were ablated by both laser wavelengths on LIBS setup. The maps in Fig. 5 show salad plants that were grown in control substrate and the Lunar simulant. The colour scale is the same in both variants within the same ablation wavelength to ensure visual comparability between variants. The same emission lines of Mg and Ca were chosen for 2D SNR maps as in the case of broccoli. The trend of higher concentrations of nutrients accumulated in plants grown in Lunar simulant has also been confirmed by salad plants. Compared to salad, in which elements are diffusely distributed throughout the leaf and not restricted to particular tissues, nutrients in broccoli are more tissue-specific and preferentially localised in the veins and midrib, with lower levels in the lamina.

The results of bioimaging by both ablation wavelengths show an increased accumulation of magnesium and calcium in the plant tissue of the Lunar regolith grown plants. This can be correlated to the composition of Lunar regolith simulant (LMS-1, Exolith Lab, USA) used for plant cultivation in this study. They are composed of a few different oxides, but among the most represented are MgO and CaO [35,36]. These oxides are not present in such high concentrations in the control substrate. That leads to an increased accumulation of these elements in plant tissue, which was confirmed by bioimaging. Duri et al. (2025) investigated the effect of LHS-1 regolith simulant (Exolith Lab, USA) on the salad (*Lactuca sativa*) plant nutrient uptake. LHS-1 simulant does not contain as much MgO as LMS-1, which led to no differences in magnesium uptake in plant tissue. However, LHS-1 simulant is significantly richer in case of CaO concentration than LMS-1 [36]. Similarly to magnesium uptake, no significant differences were observed in the case of Ca uptake [44]. In the review on the potential of regolith simulants for plant growth by Duri et al. the problem of nutrient bioavailability in Lunar regolith simulants was discussed. Plants generally take up only the bioavailable forms of elements. The elements occluded in mineral structures are not nutritionally beneficial [45]. The difference in mineralogy of LHS-1 and LMS-1 simulants might affect the plant nutrient uptake [36].

**4. Conclusion**

The need for sustainable food production for long-term space missions is inevitable. In this study, we employed laser-induced breakdown spectroscopy (LIBS) for bioimaging of plants grown in Lunar regolith simulants. For this purpose, we used two different laser ablation wavelengths, whose effectiveness for bioimaging of plant tissue was compared. We used conventional 1064 nm and compared it with 2090 nm ablation wavelength. Due to the higher SNR values, total emissivity, and the increased Mg II / Mg I ratio (ionisation degree) observed when using the 2090 nm ablation wavelength, we can assume that the plasma generated by this laser had a higher temperature. This indicates a more efficient interaction between the laser and the plant tissue compared to the 1064 nm laser.

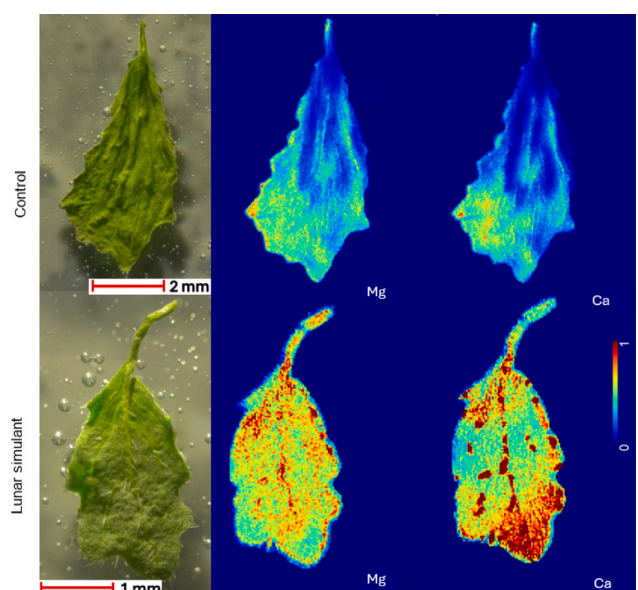

(a) 2090 nm

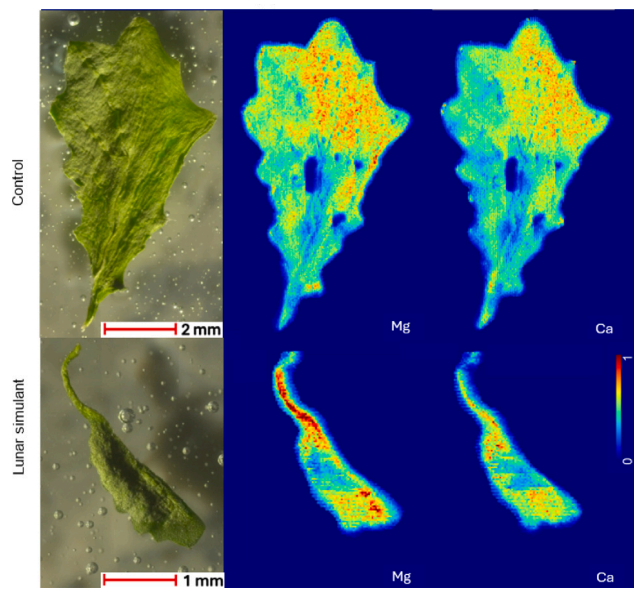

(b) 1064 nm

**Fig. 4.** Results of LIBS analysis of broccoli (*Brassica oleracea*) plants grown in lunar regolith simulant and control substrate, with two different ablation wavelengths: (a) 2090 nm and (b) 1064 nm. The heatmaps show atomic Mg I 285.31 nm and singly ionised Ca II 393.37 nm. The colour scale is the same in both variants within the same ablation wavelength.

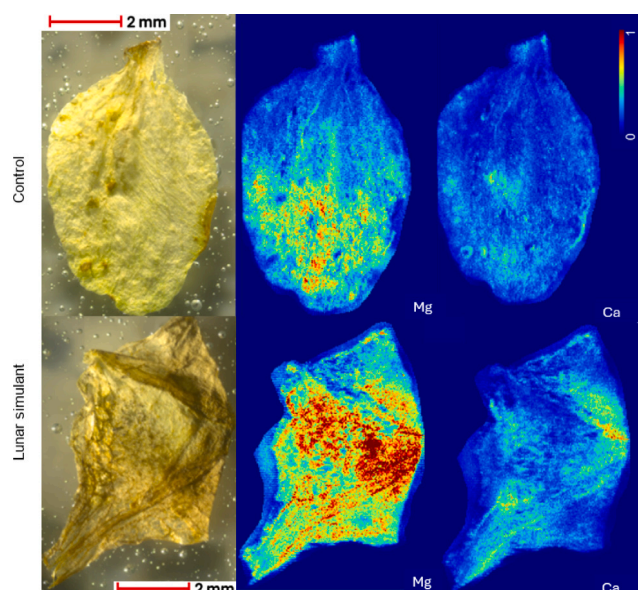

(a) 2090 nm

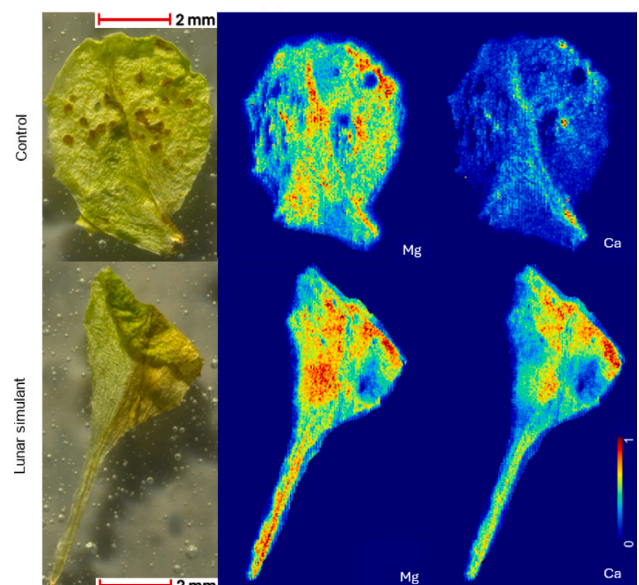

(b) 1064 nm

**Fig. 5.** Results of LIBS analysis of salad (*Lactuca sativa*) plants grown in Lunar regolith simulant and control substrate, with two different ablation wavelengths 2090 nm (a) and 1064 nm (b). The heatmaps are showing atomic Mg I 285.31 nm and singly ionised Ca II 393.37 nm. The colour scale is the same in both variants within the same ablation wavelength.

The following results of plants bioimaging showed that both plants (broccoli and salad) have accumulated higher amounts of key plant nutrients such as magnesium and calcium when grown in lunar regolith simulant, which corresponded to the composition of LMS-1 regolith simulant. These results should help us to understand better the interaction between the plant and Lunar simulant and their nutrient uptake. Moreover, when LIBS is applied for plant analysis in a Lunar environment, the 2090 nm ablation wavelength would be more suitable for analysing plant tissues. However, further research is necessary to evaluate how plants grow in lunar simulants and how they absorb nutrients.

## CRediT authorship contribution statement

**Tomáš Vozár:** Writing – review & editing, Writing – original draft, Visualization, Software, Methodology, Investigation, Data curation, Conceptualization. **Ludmila Čechová:** Writing – review & editing, Visualization, Validation, Supervision, Methodology, Investigation,

Conceptualization. **Jakub Buday:** Writing – review & editing, Validation, Supervision, Software, Methodology, Formal analysis, Data curation, Conceptualization. **Martin Füleky:** Writing – review & editing, Methodology, Investigation. **Katarína Molnárová:** Writing – review & editing, Validation, Methodology, Investigation. **Libor Lenža:** Supervision, Funding acquisition, Conceptualization. **Jiří Mužík:** Resources, Methodology, Formal analysis. **Yuya Koshiba:** Resources, Methodology. **Martin Smrž:** Supervision, Resources. **Pavel Pořízka:** Writing – review & editing, Validation, Supervision, Resources, Funding acquisition, Conceptualization. **Jozef Kaiser:** Supervision, Resources, Funding acquisition.

**Declaration of competing interest**

The authors declare that they have no known competing financial interests or personal relationships that could have appeared to influence the work reported in this paper.

**Acknowledgements**

This work was carried out with the support of grant no. 25-18588L (Grant Agency Czech Republic (The Czech Science Foundation)). JM, MS and YK acknowledge the work was co-funded by European Union and the state budget of the Czech Republic under the project LasApp CZ.02.01.01/00/22_008/0004573 and by MERIT programme 101081195 powered by Central Bohemian Innovation Center. JK acknowledges the support of grant no. FSI-S-23-8389 (Brno University of Technology). The measurements by 2090 nm laser wavelength were supported by HiLASE Centre, Czech Republic.

**Data availability**

Data will be made available on request.